\pdfoutput=1
\documentclass[letter]{IEEEtran}
\usepackage[T1]{fontenc}
\usepackage{tikz}
\usepackage{float}
\usepackage{graphicx}
\usepackage{color,soul}
\usepackage[nolist,nohyperlinks]{acronym}
\usepackage{xcolor}
\colorlet{activecolor}{black}
\colorlet{second}{black}
\usepackage{booktabs}

\usepackage[left=1.1in,top=1.2in,right=1in]{geometry}
\usepackage{cite}

\ifCLASSINFOpdf
\else
\fi
\usepackage{amsmath}
\usepackage[cmintegrals]{newtxmath}
\begin{document}
%
\title{How Functional Complexity affects the Scalability-Energy Efficiency Trade-Off of HCC WSN Clustering}
%
%
%

\author{Merim~Dzaferagic,
        Nicholas~Kaminski,
        Irene~Macaluso,
        and~Nicola~Marchetti\\
        CONNECT, Trinity College Dublin, Ireland, E-mail: \{dzaferam, kaminskn, macalusi, marchetn\}@tcd.ie}
\maketitle
\thispagestyle{empty}

\begin{abstract}
\textcolor{activecolor}{Even though clustering algorithms in Wireless Sensor Networks (WSN) are a well investigate subject, the increasing interest in the Internet of Things (IoT) and 5G technologies has precipitated the need of new ways to comprehend and overcome a new set of challenges. While studies mainly propose new algorithms and compare these algorithms based on a set of properties (e.g. energy efficiency, scalability), none of them focuses on the underlying mechanisms and organizational patterns that lead to these properties. We address this lack of understanding by applying a complex systems science approach to investigate the properties of WSNs arising from the communication patterns of the network nodes.} We represent different implementations of clustering in WSNs with a functional topology graph. Moreover, we employ a complexity metric - functional complexity ($C_F$) - to explain how local interactions give rise to the global behavior of the network. Our analysis shows that higher values of $C_F$ indicate higher scalability and lower energy efficiency.  

\end{abstract}


\IEEEpeerreviewmaketitle

\section{Introduction}\label{sec:introduction}
Clustering partitions a network of nodes into a number of smaller groups (clusters). Different approaches to clustering are available in the literature. The authors of \cite{Abbasi2007, Liu2012, Aslam2012, Banerjee2001} introduced different approaches, which involve adaptive clustering, random competition based clustering, \ac{HCC}, energy efficient hierarchical clustering, distributed clustering, \ac{LEACH}, and \ac{HEED} clustering. They also highlighted that the algorithms differ in properties like stability of the created clusters, objectives (e.g. scalability, fault-tolerance, connectivity, load balancing, redundancy elimination, rapid convergence, network lifetime), clustering criteria (e.g. identifier, position, cluster head frequency, residual energy), methodology (e.g. distributed, centralized, hybrid). \textcolor{activecolor}{Even though these algorithms are well studied, the analysis is missing an important aspect which involves the understanding of the mechanisms that lead to certain properties of these algorithms. Since the \ac{HCC} algorithm proposed in \cite{Banerjee2001} is the most popular multi-tier hierarchical clustering algorithm \cite{Abbasi2007}, in the course of this analysis we will focus on this algorithm.}

Understanding the organizational and communication characteristics of different clustering algorithms allows us to comprehend which aspects of a specific implementation lead to certain characteristics, i.e. scalability and energy efficiency. In \cite{Dzaferagic2016}, we propose a framework which allows us to represent network functions with graphs called functional topologies. Therein, we also propose a metric to calculate the functional complexity of an implementation of a network function. Here, we employ our functional framework to model an implementation for clustering in \ac{WSNs}. \textcolor{second}{Our goal is to investigate the relationship between the functional complexity and certain properties of the implementation, such as scalability and energy efficiency}. This allows us to understand the underlying mechanisms that are a product of complex interactions between functional entities.



\section{Network Functional Framework}\label{sec:Functional framework}
\textcolor{activecolor}{We employ the functional framework introduced in \cite{Dzaferagic2016} to represent the implementation of the clustering algorithm with a functional topology. The functional topology is a graph that depicts the functional connectivity between system parts, where each node represents a functional entity related to the implementation, and each link indicates interactions between nodes.}

\textcolor{activecolor}{The functional complexity metric, also proposed in \cite{Dzaferagic2016}, is based on the Shannon entropy ($H(x_n)$) as a key characteristic of any system. To describe the potential for a node to interact with other nodes we employ the Bernoulli random variable $x_n$. The probability of interactions $p(x_n=1)$ is defined as the reachability of a node $n$ ($p_r(x_n=1) = i_r^n/j$, where $i_r^n$ is the number of nodes that can reach node $n$ and $j$ is the number of nodes for the given subgraph.}


\textcolor{second}{Since we focus on one hop communication the complexity metric applied to our system turns out to be the following simplified single scale version of the general functional complexity expression introduced in \cite{Dzaferagic2016}, i.e.}
\begin{equation}\label{eq:single_scale_complexity}
C_F = \displaystyle\sum_{j = 2}^{N} | \langle I(\Lambda^j) \rangle - \dfrac{j}{N}  I(\Lambda^N)|,
\end{equation}
\textcolor{activecolor}{where $N$ is the total number of nodes in the functional topology and $\Lambda^j$ is a subgraph with $j$ nodes. The functional complexity compares the average uncertainty of interactions for a smaller subset ($\langle I(\Lambda^j) \rangle$) to the amount of information which is expected from the calculation performed on the whole system ($I(\Lambda^N)$).}

\textcolor{second}{$I(\Lambda^N)$ is the total amount of information of the whole functional topology. $I(\Lambda_k^j)$ is the uncertainty of interactions for a smaller subset, i.e. the total amount of information of the  $\mathrm{k^{th}}$ subgraph with $j$ nodes. It is calculated with equation (\ref{eq:total_amount_of_information}).}

\begin{equation}\label{eq:total_amount_of_information}
I(\Lambda_k^j) = \displaystyle\sum_{n \in \Lambda_k^j} H(x_n)
\end{equation}

$H(x_n)$ reaches its maximum if the probability of interaction with node $n$ is $p(x_n = 1) = 1/2$. As the distribution of links among nodes for a sparse graph is almost uniform, a sparse graph results in high values of $H(x_n)$. High values of $H(x_n)$ result in high values of $I(\Lambda_k^j)$. Therefore, the functional complexity is high for a sparse graph, with uniformly distributed links among nodes for subgraphs with the size $j < N$. The functional complexity is zero for a fully connected and for a disconnected graph. For more details about the functional framework and the complexity metric expressed by equation (\ref{eq:single_scale_complexity}) the reader is referred to \cite{Dzaferagic2016}. 

\begin{figure}[t]
	\centering
	\includegraphics[scale=0.4]{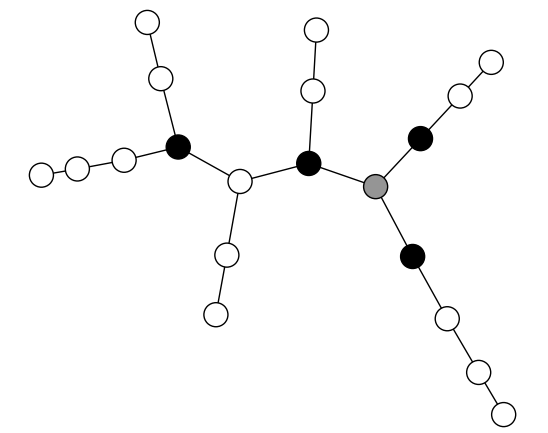}      
	\caption{The functional topology of the \ac{HCC} algorithm for a network of twenty nodes, which represents the \ac{BFS} tree created based on the physical topology. The white nodes represent ordinary nodes, the black nodes represent cluster-heads, and the gray node represents the base station. }
	\label{functional_topology_HCCA}
\end{figure}

\textcolor{activecolor}{The physical topology of our \ac{WSN} is created according to the Von Neumann neighborhood, which is basically an undirected lattice graph. Now we employ our functional framework to represent the implementation of the \ac{HCC} algorithm as a functional topology. Our goal is to investigate the influence of the interactions among nodes after the clusters are established, on the objectives of clustering algorithms (\textcolor{second}{i.e. scalability and energy efficiency}). We focus on the maintenance phase of the algorithm. After the set-up phase the nodes establish connections to their neighboring nodes according to the \ac{BFS} algorithm. Each node discovers its subtree, and exchanges information with its neighbors in the \ac{BFS} tree. We imagine a virtual decision maker entity that is moving from one node to another. At each node the decision maker entity collects information from nodes that belong to its subtree  and forwards this information to its parent. In other words, each node maintains its position in the \ac{BFS} tree and therefore the functional topology of the \ac{HCC} algorithm is equivalent to the \ac{BFS} tree created upon the physical topology (Figure \ref{functional_topology_HCCA}).}

\section{Analysis}\label{sec:Analysis}
\textcolor{activecolor}{Adding a new node to a cluster created according to the \ac{HCC} algorithm simply means that the new node gets connected to one of the existing nodes which is going to be its parent node. The new node does not need to inform all nodes in the cluster about its arrival, which simplifies the process of adding nodes to the network and represents a highly scalable implementation. According to the authors of \cite{Abbasi2007,Liu2012, Aslam2012} and \cite{Jiang2009} hierarchical approaches proved themselves to be more scalable than their non-hierarchical counterpart. The authors of \cite{Jiang2009} showed that a trade-off exists between network scalability and energy consumption in clustering schemes. Our goal is to investigate the relationship between these objectives and the functional complexity, which would allow us to analyze them together.}

\begin{figure}[t]
	\centering
	\includegraphics[scale=0.4]{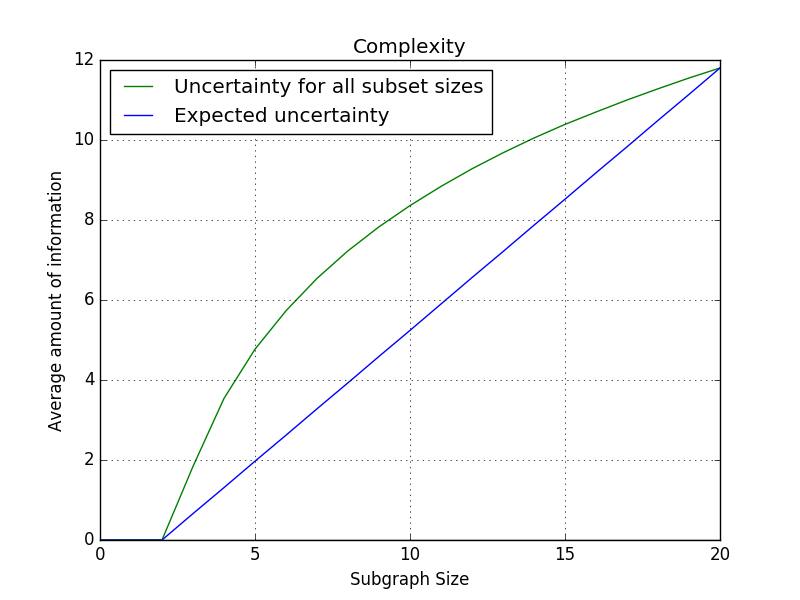}      
	\caption{Functional complexity of the \ac{HCC} algorithm; the functional complexity is the area between the green and the blue curves.}
	\label{complexity_hcc_4x5}
\end{figure}

\textcolor{activecolor}{Figure \ref{complexity_hcc_4x5} depicts the relationship between the uncertainty of interactions for all subset sizes and the amount of information which is expected from the calculation performed on the whole system for the \ac{HCC} algorithm, i.e. the functional complexity of \ac{HCC}. As shown in Figure \ref{functional_topology_HCCA} the functional topology of the \ac{HCC} algorithm has sparse intra and inter cluster connections. As the links in the functional topology represent functional dependencies between nodes, a sparse connectivity pattern indicates weak dependencies which result in high scalability. This follows from the fact that in order to add a node to the network, the new node \textcolor{second}{only needs} to establish a connection (send a message) to one of the nodes in the topology and to declare this node as its parent. The functional complexity of the \ac{HCC} algorithm is 38.31. As the authors of \cite{Abbasi2007,Liu2012, Aslam2012} and \cite{Jiang2009} agree that the communication between the cluster-head and the base station consumes most energy, an approximation of the energy efficiency could be calculated as the ratio between the average number of intra-cluster connections and the \textcolor{second}{total number} of links between the base station and each cluster-head in the functional topology.The energy efficiency of the \ac{HCC} algorithm is 0.61 which is very low.}

\section{Conclusion}\label{sec:conclulsion}
Our analysis confirms the observation made by the authors of \cite{Jiang2009}, which highlights the trade-off between scalability and energy efficiency. With our complex systems science approach we showed that these aspects of different clustering algorithms can be analyzed together by analyzing the functional complexity of the specific implementation. Increasing values of $C_F$ lead to the increase of scalability and the decrease of energy efficiency. \textcolor{second}{Our ongoing work also confirms this trend, by analyzing other WSN clustering algorithms.}

\section*{Acknowledgment}
This material is based upon works supported by the Science Foundation Ireland under the Grant No. 13/RC/2077.

\ifCLASSOPTIONcaptionsoff
  \newpage
\fi


\bibliographystyle{templates/IEEEtran}  
\bibliography{clustering_in_wsn}



\begin{acronym}
	\acro{RNC}{Radio Network Controler}
	\acro{nodeB}{UMTS base transciever station}
	\acro{VLR}{Visitor Location Register} 
	\acro{HLR}{Home Location Register}
	\acro{SGSN}{Serving GPRS Support Node}
	\acro{MSC}{Mobile Switching Center}
	\acro{SMS}{Short Message Service}
	\acro{IoT}{Internet of Things}
	\acro{AP}{Access Points}
	\acro{WSNs}{Wireless Sensor Networks}
	\acro{WSN}{Wireless Sensor Network}
	\acro{IoT-GSI}{Global Standards Initiative on Internet of Things}
	\acro{DCA}{Distributed Clustering Algorithm}
	\acro{BFS}{Breadth-First Search}
	\acro{LEACH}{Low Energy Adaptive Clustering Hierarchy}
	\acro{HCC}{Hierarchical Control Clustering}
	\acro{HEED}{Hybrid Energy-Efficient Distributed}
\end{acronym}

\end{document}